\documentclass[prd,twocolumn,showpacs,preprintnumbers,amsmath,amssymb,superscriptaddress,floatfix,nofootinbib]{revtex4-2}

\usepackage{graphicx}
\usepackage{bm}
\usepackage{amsmath}
\usepackage{amsfonts}
\usepackage{amssymb}
\usepackage{color}
\usepackage{subfigure}

\usepackage[colorlinks, citecolor=blue,anchorcolor=red,menucolor=red, linkcolor=red,filecolor=red,runcolor=red,urlcolor=blue,frenchlinks=red]{hyperref}

\begin{document}
\title{Theoretical study of  the open-flavored tetraquark $T_{c\bar{s}}(2900)$ in the process $\bar{B}_s^0 \to K^0 D^0 \pi^0$}
\author{Wen-Tao Lyu}\email{wentaolyu@gs.zzu.edu.cn}
\affiliation{School of Physics, Zhengzhou University, Zhengzhou 450001, China}

\author{Man-Yu Duan}\email{duanmy@seu.edu.cn}
\affiliation{School of Physics, Southeast University, Nanjing 210094, China}
\affiliation{School of Physics, Zhengzhou University, Zhengzhou 450001, China}
\affiliation{Departamento de Física Teórica and IFIC, Centro Mixto Universidad de Valencia-CSIC Institutos de Investigación de Paterna, 46071 Valencia, Spain}

\author{Xiaoyu Wang}\email{xiaoyuwang@zzu.edu.cn}
\affiliation{School of Physics, Zhengzhou University, Zhengzhou 450001, China}

\author{Dian-Yong Chen}\email{chendy@seu.edu.cn}
\affiliation{School of Physics, Southeast University, Nanjing 210094, China}
\affiliation{Lanzhou Center for Theoretical Physics, Lanzhou University, Lanzhou 730000,China}

\author{En Wang}\email{wangen@zzu.edu.cn}
\affiliation{School of Physics, Zhengzhou University, Zhengzhou 450001, China}

\begin{abstract}
Recently, the LHCb Collaboration has measured two decay processes $B^0\to\bar{D}^0D_s^+\pi^-$ and $B^+\to D^-D_s^+\pi^+$ related to isospin symmetry, where two new open-flavored tetraquark states $T_{c\bar{s}}(2900)^0$ and $T_{c\bar{s}}(2900)^{++}$ that belong to an isospin triplet were observed in the $D_s^+\pi^-$ and $D_s^+\pi^+$ invariant mass distributions.
In this work, we have investigated the validity of the process $\bar{B}_s^0\to K^0D^0\pi^0$ as the promising process to confirm the existence of $T_{c\bar{s}}(2900)^0$ resonance.
Taking into account the tetraquark state $T_{c\bar{s}}(2900)$, as well as intermediate resonances $K^*(892)$, $K_0^*(1430)$, and $K_2^*(1430)$, it has been shown that a clear peak of the open-flavored tetraquark $T_{c\bar{s}}(2900)$ appears in the $K^0D^0$ invariant mass distribution of the process $\bar{B}_s^0\to K^0D^0\pi^0$, which could be tested by future experiments.
\end{abstract}
	
	\date{\today}
	
	\maketitle
	
\section{Introduction}\label{sec1}
	
In 2020, the LHCb Collaboration has analyzed the relevant data of the process $B^+\to D^+D^-K^+$, and observed two charm-strange resonances $X_0(2900)$ and $X_1(2900)$ with masses around 2900~MeV in the $D^-K^+$ invariant mass distribution~\cite{LHCb:2020bls,LHCb:2020pxc}. Since they contain four different flavor quarks ($\bar{c}du\bar{s}$) and cannot be described within conventional quark models, these two resonances have attracted a lot of attention. Some interpretations of the two exotic resonances with open flavors have been proposed, such as compact tetraquark, molecular states, and sum rules technique (among others, see Ref.~\cite{Dai:2022qwh}). Since the mass of $X_0(2900)$ is close to the $\bar{D}^*K^*$ threshold, the $\bar{D}^*K^*$ molecular explanation is more popular~\cite{Liu:2020nil, Huang:2020ptc,Hu:2020mxp,Kong:2021ohg,Wang:2021lwy,Xiao:2020ltm,Lin:2022eau,Chen:2020eyu,Liu:2020orv, Chen:2020aos,Chen:2021erj,Albuquerque:2020ugi,Xiao:2020ltm}. In particular, using the method of QCD sum rules, the authors in Refs.~\cite{Chen:2020aos,Chen:2021erj,Albuquerque:2020ugi} have interpreted $X_0(2900)$ as the $S$-wave $\bar{D}^{*}K^{*}$ molecular state of $J^P=0^+$, and suggested that $X_1(2900)$ can be interpreted as the $P$-wave $\bar{c}\bar{s}ud$ compact tetraquark state of $J^P=1^-$. 
Meanwhile, $X_0(2900)$ could be regarded as a $D^*\bar{K}^*$ molecule state within the one-boson exchange model~\cite{Liu:2020nil}.  Recently, both $X_0(2900)$ and $X_1(2900)$ (renamed $T_{\bar c\bar s 0}(2870)$ and $T_{\bar c\bar s 1}(2900)$) were confirmed in the process $B^+\to D^{*+}D^-K^+$ by the LHCb Collaboration~\cite{LHCb:2024vfz}.

Two years after the observation of $X_{0,1}(2900)$, another doubly charged tetraquark candidate $T_{c\bar{s}}(2900)^{++}$ and its neutral partner $T_{c\bar{s}}(2900)^{0}$ that have charm and strangeness, were observed by the LHCb Collaboration in the $D_s^+\pi^+$ and $D_s^+\pi^-$ invariant mass distributions of the processes $B^+\to D^-D_s^+\pi^+$ and $B^0\to\bar{D}^0D_s^+\pi^-$, respectively~\cite{LHCb:2022sfr,LHCb:2022lzp}. The quantum numbers of their spin parity are determined
to be $0^+$ with a significance of about 7.5$\sigma$ with respect to the $1^-$ hypothesis. The quark components of these two states should be $c\bar{s}\bar{u}d$ and $c\bar{s}u\bar{d}$, and their masses and widths are determined to be
	\begin{equation}
		\begin{aligned}
			& M_{T_{c \bar{s}}(2900)^0}=(2892 \pm 14 \pm 15)~\mathrm{MeV} \text { , } \\
			& \Gamma_{T_{c \bar{s}}(2900)^0}=(119 \pm 26 \pm 12)~\mathrm{MeV} \text { , } \\
			& M_{T_{c \bar{s}}(2900)^{++}}=(2921 \pm 17 \pm 19)~\mathrm{MeV} \text { , } \\
			& \Gamma_{T_{c \bar{s}}(2900)^{++}}=(137 \pm 32 \pm 17)~\mathrm{MeV} \text { . } 
		\end{aligned}
	\end{equation}
The significance is found to be $8.0\sigma$ for the $T_{c\bar{s}}(2900)^0$ state and $6.5\sigma$ for the $T_{c\bar{s}}(2900)^{++}$ state including systematic uncertainties. 
The mass and width differences between $T_{c\bar{s}}(2900)^{++}$ and $T_{c\bar{s}}(2900)^{0}$ are evaluated as~\cite{LHCb:2022sfr,LHCb:2022lzp} 
\begin{align}
    & \Delta M =(28\pm20\pm12)~\rm{MeV,}\nonumber\\
    & \Delta \Gamma =(15\pm39\pm16)~\rm{MeV,}
\end{align}
which can be recognized as consistent with zero. Thus, these two states should be two of the isospin triplets with four different quark flavors.
	
It is worth mentioning that such $C=1, S=1$ bound states were predicted more than ten years ago utilizing the hidden-gauge formalism in a coupled channel unitary approach in Ref.~\cite{Molina:2010tx}.
After the observation of $T_{c\bar{s}}(2900)^{++}$ and $T_{c\bar{s}}(2900)^0$, there are several interpretations about their structure. 
The proximity of the $D^*K^*$ and the $D_s^*\rho$ thresholds to the mass of $T_{c\bar{s}}(2900)$ suggests that these two-hadron channels could play an important role in the dynamics of the $T_{c\bar{s}}(2900)$ states, hinting at a hadronic molecular interpretation of the two states~\cite{Yue:2022mnf,Chen:2022svh}. 
For instance, Ref.~\cite{Agaev:2022duz} argues that $T_{c\bar{s}}(2900)^{++}$ and $T_{c\bar{s}}(2900)^0$ can be modeled as molecules $D_s^{*+}\rho^+$ and $D_s^{*+}\rho^-$, respectively, using the two-point sum rule method. 
Recently, within the local hidden symmetry formalism, $T_{c\bar{s}}(2900)$ could be explained as a $D^*_s\rho-D^*K^*$ bound/virtual state in Ref.~\cite{Duan:2023lcj}. We have also proposed to search for the open-flavored tetraquark $T_{c\bar{s}}(2900)$ in the processes $B^+\to D^+D^-K^+$~\cite{Duan:2023qsg}, $B^-\to D_s^+K^-\pi^-$~\cite{Lyu:2023jos}, and $\Lambda_b\to D^0K^0\Lambda$~\cite{Lyu:2023aqn} by assuming $T_{c\bar{s}}(2900)$ as a $D^*K^*$ molecular state. Meanwhile, its spin partner with $J^P=2^+$ predicted in Refs.~\cite{Molina:2022jcd,Duan:2023lcj} is suggested to be studied in the processes $B\to D^{*-}D^+K^+$ and  $\Lambda_b\to \Sigma_c^{++}D^{-}K^{-}$~\cite{Lyu:2024zdo,Song:2024dan}.

In addition, the compact tetraquark interpretation of the $T_{c\bar{s}}(2900)$ with quark contents $[c\bar{s}\bar{u}d]$ and $[c\bar{s}u\bar{d}]$ is studied in Refs.~\cite{Yang:2023evp, Lian:2023cgs, Jiang:2023rcn, Liu:2022hbk, Dmitrasinovic:2023eei, Ortega:2023azl}, while the $T_{c\bar{s}}(2900)$ structure is also proposed to be a threshold cusp effect from the interaction between the $D^*K^*$ and $D^*_s\rho$ channels~\cite{Molina:2022jcd} or the kinetic effect from a triangle singularity~\cite{Ge:2022dsp}. Thus, searching for $T_{c\bar{s}}(2900)$ in other processes is crucial to excluding the interpretation of kinetic effects, and more experimental information could be helpful to explore its structure and deepen our understanding of the hadron-hadron interaction.
	
Recently, Ref.~\cite{Qin:2022nof} has studied the production processes of hidden-charm and open-charm tetraquarks in $B$ decays by analyzing their topological amplitudes, and suggested the experimental searches for $\bar{B}_s^0\to T_{c\bar{s}}(2900)\pi^0 \to K^0D^0\pi^0$ with priority. Taking into account that $T_{c\bar{s}}(2900)$ is predicted to strongly couple to the $DK$ channel~\cite{Duan:2023lcj,Yue:2022mnf,Lian:2023cgs}, we would like to propose to search for this state in $\bar{B}_s^0\to K^0D^0\pi^0$, which has not been measured experimentally yet. However, the process $B_s^0\to \bar{D}^0K^-\pi^+$ has been observed by the LHCb Collaboration~\cite{LHCb:2013svv,LHCb:2014ioa,Craik:2015nga,LHCb:2013vvh,LHCb:2011yev}. And we can find the branching fraction $\mathcal{B}(B_s^0\to \bar{D}^0K^-\pi^+)=(1.04\pm0.13)\times 10^{-3}$~\cite{LHCb:2013svv,ParticleDataGroup:2022pth}. Since both the tree-diagram amplitudes of processes $\bar{B}_s^0\to K^0D^0\pi^0$ and $B_s^0\to \bar{D}^0K^-\pi^+$ are proportional to the Cabbibo-Kobayashi-Maskawa (CKM) matrix elements $V_{cb}V_{ud}$, it is expected that the branching fractions of those processes should be of the same order of magnitude, and the process $\bar{B}_s^0\to K^0D^0\pi^0$ is expected to be observed by the LHCb Collaboration. Thus, in this work, we would like to investigate the process $\bar{B}_s^0\to K^0D^0\pi^0$ to show the possible evidence of $T_{c\bar{s}}(2900)$ in the $D^0K^0$ invariant mass distribution, rather than the process $\bar{B}^0_s\to D^0K^+\pi^-$, since the $D^0K^+$ system could not distinguish the isospin $I=0$ and $I=1$. Here we focus on the decay process $\bar{B}_s^0\to K^0D^0\pi^0$ by considering the $D_s^{*+}\rho^-$ and $D^{*0}K^{*0}$ final-state interaction to generate the signal of $T_{c\bar{s}}(2900)^0$. In addition, we will also consider contributions from the intermediate states $K^*(892)$, $K_0^*(1430)$, and $K_2^*(1430)$.
	
This paper is organized as follows. In Sect.~\ref{sec2}, we present the theoretical formalism of the process $\bar{B}_s^0\to K^0D^0\pi^0$. The numerical results are presented in Sect.~\ref{sec3}. Finally, in the last section, we present a brief summary.
	
\section{Formalism}\label{sec2}
	
In this section, we will introduce the theoretical formalism for the process $\bar{B}_s^0\to K^0D^0\pi^0$. Firstly, the mechanism for this process via the intermediate state $T_{c\bar{s}}(2900)$ is given in Sect.~\ref{sec2a}, and then the mechanism via the intermediate excited $K^*$ states is given in Sec.~\ref{sec2b}. Finally, we give the formalism for the invariant mass distributions for this process in Sect.~\ref{sec2e}. 
	
\subsection{$T_{c\bar{s}}(2900)$ in the $\bar{B}_s^0\to K^0D^0\pi^0$} \label{sec2a}
	
\begin{figure}[htbp]
		\subfigure[]{\includegraphics[scale=0.63]{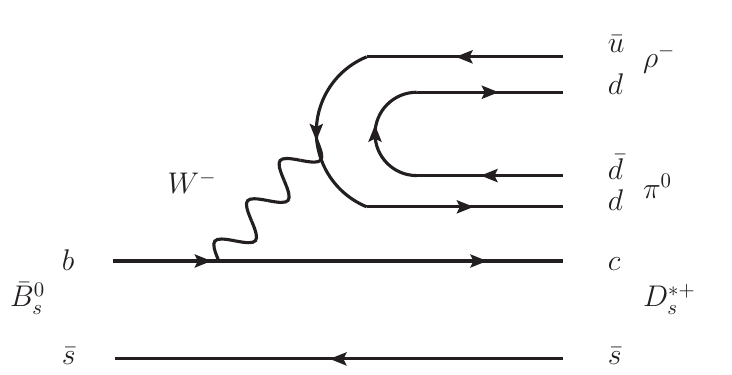}}	
		\subfigure[]{\includegraphics[scale=0.63]{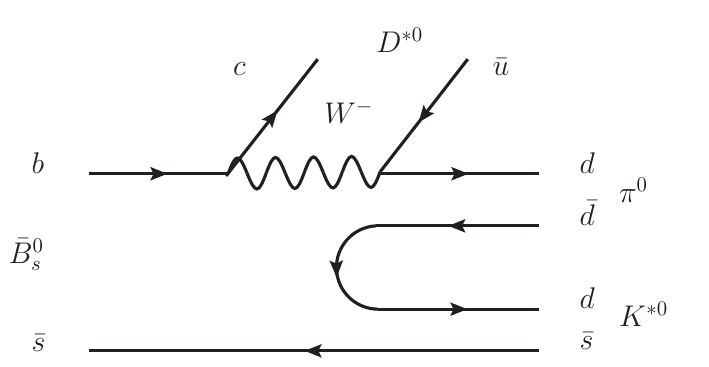}}
		\caption{The $\bar{B}_s^0$ weak decay at the microscopic quark picture. (a) The $\bar{B}_s^0\to\rho^-\pi^0D_s^{*+}$ decay via the mechanism of the external $W^-$ emission, (b) the $\bar{B}_s^0\to D^{*0}\pi^0K^{*0}$ decay via the mechanism of the internal $W^-$ emission.}\label{fig:Tcs-quark}
\end{figure}

Since $T_{c\bar{s}}(2900)$ could be explained as the $D^*_s\rho-D^*K^*$ bound/virtual state in the coupled-channel approach~\cite{Duan:2023lcj}, it is necessary to produce $D^*_s\rho$ or $D^*K^*$ through the mechanisms of the external and the internal $W^-$ emission, as depicted in Figs.~\ref{fig:Tcs-quark}(a) and \ref{fig:Tcs-quark}(b), respectively.
	
For the mechanism of external $W^-$ emission, the $b$ quark of the initial $\bar{B}_s^0$ weakly transits into a $W^-$ boson and a $c$ quark, followed by the $W^-$ decaying into the $\bar{u}d$ pair. As depicted in Fig.~\ref{fig:Tcs-quark}(a), the $\bar{u}d$ pair from the $W^-$ boson decay, together with the $\bar{d}d$ pair created from the vacuum, will hadronize into $\rho^-$ and $\pi^0$, and the $c$ and $\bar{s}$ will hadronize into $D_s^{*+}$ meson, as follows,
	\begin{eqnarray}\label{H1}
		\left| \bar{B}_s^0 \right\rangle &=& b\bar{s} \nonumber \\
		&\Rightarrow & W^- c \bar{s} \nonumber \\
		&\Rightarrow & \bar{u}  \left( d\bar{d}\right) dc\bar{s} \nonumber \\
		&\Rightarrow &-\frac{1}{\sqrt{2}} \rho^-\pi^0D_s^{*+}.
	\end{eqnarray}

On the other hand, we can also produce the $D^{*0} K^{*0}$ via the mechanism of the internal $W^-$ emission, as depicted in Fig.~\ref{fig:Tcs-quark}(b) as follows,
\begin{eqnarray}\label{H2}
		\left| \bar{B}_s^0 \right\rangle &=& b\bar{s} \nonumber \\
		&\Rightarrow & cW^- \bar{s} \nonumber \\
		&\Rightarrow & c\bar{u} d \left( \bar{d}d\right) \bar{s} \nonumber \\
		&\Rightarrow &-\frac{1}{\sqrt{2}}  D^{*0}\pi^0K^{*0} .
\end{eqnarray}

\begin{figure}[htbp]
		\subfigure[]{\includegraphics[scale=0.63]{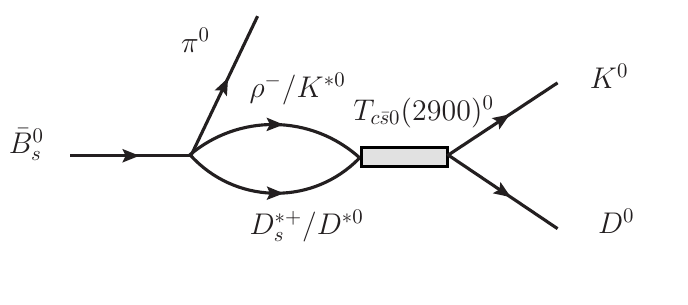}}	
		\subfigure[]{\includegraphics[scale=0.5]{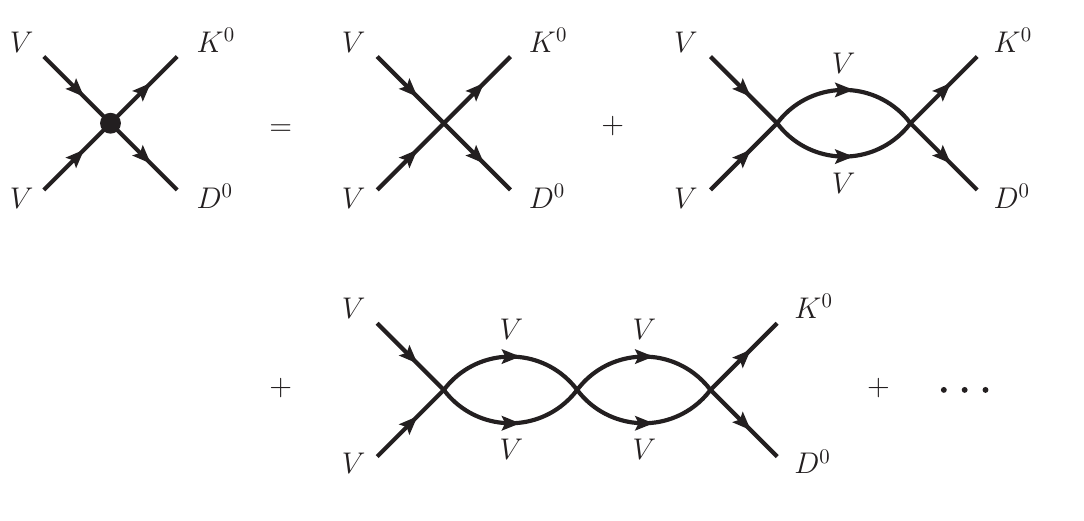}}
		\caption{(a) $\bar{B}_s^0\to K^0D^0\pi^0$ decay via the intermediate state $T_{c\bar{s}}(2900)$ produced by the interactions of $D_s^{*+}\rho^-$ and $D^{*0}K^{*0}$. (b) The re-scattering process within a trivial summation.}\label{fig:Tcs-loop}
\end{figure}

Now, the systems of $D^{*+}_s\rho^-$ and $D^{*0} K^{*0}$ could undergo $S$-wave interactions to generate the $T_{c\bar{s}}(2900)$ state, followed by decaying into the final states $D^0K^0$, as depicted in Fig.~\ref{fig:Tcs-loop}(a). In this case, we could obtain the unitary amplitude for the final state interactions following a trivial summation as depicted in Fig.~\ref{fig:Tcs-loop}(b), 
	\begin{eqnarray}
			&&  \tilde{G}_{i} V'_{i\to D^0K^0} + \tilde{G}_{i} V_{i\to j} \tilde{G}_{j} V'_{j\to D^0K^0} +... \nonumber \\
			&=& \tilde{G}[1-V\tilde{G}]^{-1}V',
	\end{eqnarray}
where $\tilde{G}$ is the loop functions of the two-meson system, $V'_{i\to D^0K^0}$ is the transition amplitude between the vector-vector channel and the $D^0K^0$ channel, and $V_{i\to j}$ are the transition potential between vector-vector channels~\cite{Molina:2022jcd},
    \begin{equation}
			{V}=\left(\begin{array}{cc}
				0.11g^2 & -6.8g^2 \\
				-6.8g^2& 0 
			\end{array}\right),
	\end{equation}
with $g=M_V/2f$, and $i,j=1,2$ corresponding to $D^{*0}K^{*0}$, $D_s^*\rho$. Here we take the $M_V=M_\rho$ and the pion decay constant $f=93$~MeV~\cite{Molina:2010tx}.
	
By defining the $\tilde{t}=[1-V\tilde{G}]^{-1}$, we could write the amplitude for Fig.~\ref{fig:Tcs-loop}(a),
    \begin{equation}\label{ttcs}
		\begin{aligned}
			\mathcal{T}^{T_{c\bar{s}}} = &  \frac{-1}{\sqrt{2}} V_p \left[\tilde{G}_{D^{*0}K^{*0}}\tilde{t}_{D^{*0}K^{*0}\to D_s^{*+}\rho^-}V'_{D_s^{*+}\rho^-\to D^0K^0 }\right.  \\			
			& \left.+ \tilde{G}_{D^{*0}K^{*0}}\tilde{t}_{D^{*0}K^{*0}\to D^{*0}K^{*0}}V'_{D^{*0}K^{*0}\to D^0K^0}\right. \\			
			&\left.+ C\times \tilde{G}_{D_s^{*+}\rho^-}\tilde{t}_{D_s^{*+}\rho^-\to D_s^{*+}\rho^-}V'_{D_s^{*+}\rho^-\to D^0K^0}\right. \\			
			&\left.+C\times \tilde{G}_{D_s^{*+}\rho^-}\tilde{t}_{D_s^{*+}\rho^-\to D^{*0}K^{*0}}V'_{D^{*0}K^{*0}\to D^0K^0} \right],  
		\end{aligned}
	\end{equation}
where the parameter $V_{p}$ contains all dynamical factors of the production vertex of Fig.~\ref{fig:Tcs-quark}(b). Since in this work we mainly focus on the final state interactions of this process, and $\bar{B}_s^0$ could decay into the $\pi^0$ and $K^{*0} D^{*0}$ system with $J^P=0^+$ in the $S$-wave, we assume $V_{p}$ to be constant, as done in Refs.~\cite{Wei:2021usz,Liu:2020ajv,Wang:2015pcn,Lu:2016roh}. In addition, for the $W^-$ external emission of Fig.~\ref{fig:Tcs-quark}(a), the $\bar{u}d$ pair from the $W^-$ boson, together with the $\bar{d}d$ pair created from the vacuum, can form the color singlet $\rho^-$ and $\pi^0$, thus the $\bar{u}$ and $d$ quarks could have three choices of color. However, for the $W^-$ internal emission of Fig.~\ref{fig:Tcs-quark}(b), the $c$ quark from $b$ quark has the same color as the $b$ quark, and the $\bar{u}$ quark with the anti-color will hadronize to the color singlet $D^{*0}$, which implies that all the quarks in the final states have the fixed colors. Thus, we introduce a color factor $C=3$ to account for the relative weight of the $W^-$ external emission of Fig.~\ref{fig:Tcs-quark}(a) with respect to the $W^-$ internal mission of Fig.~\ref{fig:Tcs-quark}(b) in the case of color number $N_C=3$~\cite{Duan:2020vye,Zhang:2020rqr,Dai:2018nmw}. Within the dimensional regularization method, the loop function $G_{i}$ can be written as
	\begin{equation}\label{Gb}
		\begin{aligned}
			G_{i} & =i \int \frac{d^4 q}{(2 \pi)^4} \frac{1}{(p-q)^2-m_2^2+i \epsilon} \frac{1}{q^2-m_1^2+i \epsilon} \\
			& =\frac{1}{16 \pi^2}\left\{a_i+\ln \frac{m_1^2}{\mu^2}+\frac{s+m_2^2-m_1^2}{2 s} \ln \frac{m_2^2}{m_1^2}\right. \\
			& +\frac{|\vec{q}\,|}{\sqrt{s}}\left[\ln \left(s-\left(m_2^2-m_1^2\right)+2 |\vec{q}\,| \sqrt{s}\right)\right. \\
			& +\ln \left(s+\left(m_2^2-m_1^2\right)+2 |\vec{q}\,| \sqrt{s}\right) \\
			& -\ln \left(-s+\left(m_2^2-m_1^2\right)+2 |\vec{q}\,| \sqrt{s}\right) \\
			& \left.\left.-\ln \left(-s-\left(m_2^2-m_1^2\right)+2 |\vec{q}\,| \sqrt{s}\right)\right]\right\},
		\end{aligned}
	\end{equation}
where $m_{1,2}$ are the meson masses of the channels $D_s^{*+}\rho^-$ and $D^{*0}K^{*0}$. Here we take $\mu=1300$~MeV and $a=-1.474$ for both channels, which are the same as those in the study of the $D_s^{*+}\rho^-$ interaction in Refs.~\cite{Molina:2020hde,Molina:2022jcd,Duan:2023qsg}. In addition, we consider the decay widths of the $\rho$ and $K^*$ mesons by means of the convolution of the two-meson loop function with an energy dependent width~\cite{Ding:2023eps,Ding:2024lqk},
	\begin{equation}
		\begin{aligned}
			\tilde{G}(s)=&\frac{1}{N}\int^{M^2_{\rm max}}_ {M^2_{\rm min}}d\tilde{m}_1^2\left(-\frac{1}{\pi}\right)G(s,\tilde{m}_1^2,M_2^2) \\
			&\times {\rm Im}\left[\dfrac{1}{\tilde{m}_1^2-M_1^2+i\tilde{\Gamma}(\tilde{m})\tilde{m}_1}\right],
		\end{aligned}
	\end{equation}
with
	\begin{equation}
		N=\int^{M^2_{\rm max}}_ {M^2_{\rm min}}d\tilde{m}_1^2\left(-\frac{1}{\pi}\right){\rm Im}\left[\dfrac{1}{\tilde{m}_1^2-M_1^2+i\tilde{\Gamma}(\tilde{m})\tilde{m}_1}\right],
	\end{equation}
where $M_1$ is the nominal mass of the vector meson ($\rho$ or $K^*$), $M_{\rm min}=M_1-3.5\Gamma_0$, $M_{\rm max}=M_1+3.5\Gamma_0$~\cite{Molina:2022jcd}, with $\Gamma_0$ the resonance width at the nominal mass of the $\rho$ and $K^*$ mesons, and 
	\begin{equation}
		\tilde{\Gamma}(\tilde{m})=\Gamma_0\frac{q_{\rm off}^3}{q_{\rm on}^3}\theta(\tilde{m}-m_a-m_b),
	\end{equation}
with
	\begin{equation}
		q_{\rm off}=\dfrac{\lambda^{1/2}(\tilde{m}^2,m_a^2,m_b^2)}{2\tilde{m}},~~
		q_{\rm on}=\dfrac{\lambda^{1/2}(M_1^2,m_a^2,m_b^2)}{2M_1},
	\end{equation}
where $m_a=m_b=m_{\pi}$ for the $\rho$, and $m_a=m_{K}$, $m_b=m_{\pi}$ for the $K^*$. 
	
The transition amplitudes $V'_{i\to D^0K^0}$ of Eq.~(\ref{ttcs}) are given by
	\begin{equation}\label{t21} 
		V'_{D_s^{*+}\rho^-\to D^0K^0}=\dfrac{g_{T_{c\bar{s}},D_s^{*+}\rho^-}g_{T_{c\bar{s}},D^0K^0}}{M_{D^0K^0}^2-m_{T_{c\bar{s}}}^2+im_{T_{c\bar{s}}}\Gamma_{T_{c\bar{s}}}},
	\end{equation}
	\begin{equation}\label{t22}
		V'_{D^{*0}K^{*0}\to D^0K^0}=\dfrac{g_{T_{c\bar{s}},D^{*0}K^{*0}}g_{T_{c\bar{s}},D^0K^0}}{M_{D^0K^0}^2-m_{T_{c\bar{s}}}^2+im_{T_{c\bar{s}}}\Gamma_{T_{c\bar{s}}}},
	\end{equation}
where the $m_{T_{c\bar{s}}}=2892$~MeV and $\Gamma_{T_{c\bar{s}}}=119$~MeV are given by Refs.~\cite{LHCb:2022sfr,LHCb:2022lzp}. The constant $g_{T_{c\bar{s}},D^{*0}K^{*0}}$ corresponds to the coupling between $T_{c\bar{s}}(2900)$ and its components $D^{*0}K^{*0}$, which could be related to the binding energy by~\cite{Weinberg:1965zz,Baru:2003qq,Wu:2023fyh,Albaladejo:2022sux},
	\begin{equation}\label{g_D^{*0}K^{*0}}
		g_{T_{c\bar{s}},D^{*}K^{*}}^2=16\pi(m_{D^*}+m_{K^*})^2\tilde{\lambda}^2\sqrt{\frac{2\Delta E}{\mu}},
	\end{equation}
where $\tilde{\lambda}=1$ gives the probability of finding the molecular component in the physical states, $\Delta E=m_{D^*}+m_{K^*}-m_{T_{c\bar{s}}}$ denotes the binding energy, and $\mu=m_{D^*}m_{K^*}/(m_{D^*}+m_{K^*})$ is the reduced mass. 
	
The value of the coupling constant $g_{T_{c\bar{s}},\rho^-D_s^{*+}}$ could be obtained from the partial width of $T_{c\bar{s}}(2900)\to\rho^-D_s^{*+}$, which could be expressed as follows,
	\begin{equation}\label{gg2}
		\Gamma_{T_{c\bar{s}}\to \rho^-D_s^{*+}} =\frac{3}{8\pi}\frac{1}{m_{T_{c\bar{s}}}^2}|g_{T_{c\bar{s}}, \rho^-D_s^{*+}}|^2 |\vec{q}_{\rho}| ,
	\end{equation}
where $\vec{q}_{\rho}$ is the three-momentum of the $\rho^-$ in the $T_{c\bar{s}}(2900)$ rest frame,
	\begin{equation}
		|\vec{q}_{\rho}|=\dfrac{\lambda^{1/2}(m_{T_{c\bar{s}}}^2,m_{D_s^{*+}}^2,m_{\rho^-}^2)}{2m_{T_{c\bar{s}}}}
	\end{equation}
with the K$\ddot{a}$llen function $\lambda(x,y,z)=x^2+y^2+z^2-2xy-2yz-2zx$. And similarly we can get the $g_{T_{c\bar{s}},D^0K^0}$ coupling from the partial width of $T_{c\bar{s}}(2900)\to D^0K^0$,
	\begin{equation}
		\Gamma_{T_{c\bar{s}}\to D^0K^0} =\frac{1}{8\pi}\frac{1}{m_{T_{c\bar{s}}}^2}|g_{T_{c\bar{s}},D^0K^0}|^2 |\vec{q}_{K^0}| 
	\end{equation}
with
	\begin{equation}
		|\vec{q}_{K^0}|=\dfrac{\lambda^{1/2}(m_{T_{c\bar{s}}}^2,m_{D^{0}}^2,m_{K^0}^2)}{2m_{T_{c\bar{s}}}}.
	\end{equation}


In Ref.~\cite{Yue:2022mnf}, the authors have estimated the decay widths of $T_{c\bar{s}}(2900)$, which are $\Gamma_{T_{c\bar{s}}\to \rho^-D_s^{*+}}$=($2.96 \sim 5.3$)~MeV and $\Gamma_{T_{c\bar{s}}\to D^0K^0}=$($52.6 \sim 101.7$)~MeV. It should be found that the signal of the $T_{c\bar{s}}$ will strongly depend on the value  of $\Gamma_{T_{c\bar{s}}\to D^0K^0}$. Indeed, the width $\Gamma_{T_{c\bar{s}}\to D^0K^0}$ is predicted in many theoretical works~\cite{Duan:2023qsg,Lian:2023cgs,Huang:2023fvj}, which indicates that the width $\Gamma_{T_{c\bar{s}}\to D^0K^0}$ should be $50\sim100$~MeV. In this work, we take the center values $\Gamma_{T_{c\bar{s}}\to \rho^-D_s^{*+}}=4.13$~MeV and $\Gamma_{T_{c\bar{s}}\to D^0K^0}=77.15$~MeV, and obtain the coupling constants $g_{T_{c\bar{s}},D_s^{*+}\rho^-}=2007$~MeV, $g_{T_{c\bar{s}},D^0K^0}=4697$~MeV, and $g_{T_{c\bar{s}},D^{*0}K^{*0}}=8809$~MeV, respectively.

\subsection{$K^*(892)$, $K_0^*(1430)$, and $K_2^*(1430)$ in the $\bar{B}_s^0\to K^0D^0\pi^0$} \label{sec2b}
	
	\begin{figure}[htbp]
		
		\includegraphics[scale=0.65]{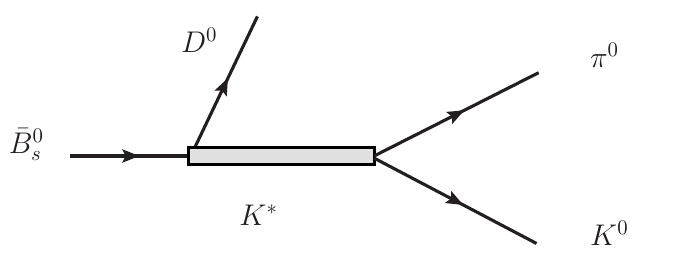}
		
		\caption{The contribution from the intermediate $K^*$.}\label{fig:Kstar}
	\end{figure}

\begin{figure}[htbp]
	\centering
	\includegraphics[scale=0.40]{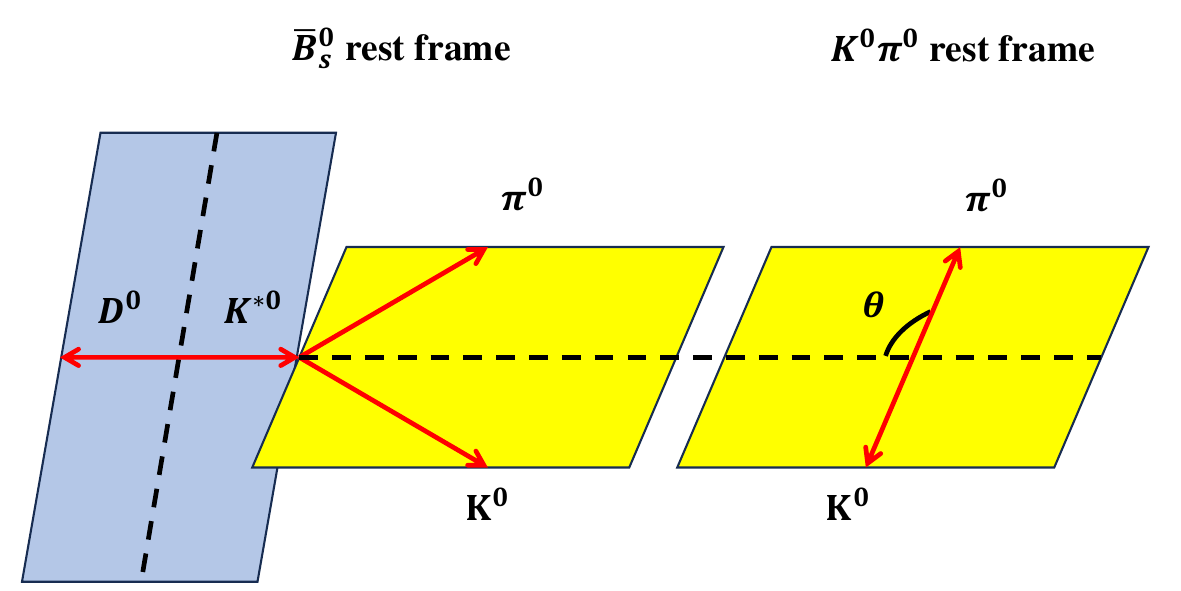}
	\caption{The angle $\theta$ between $\pi^0$ and $D^0$ in the center of mass frame of the $K^0\pi^0$ rest frame.}\label{fig:angle-diagram}
\end{figure}

\begin{table}[htpb]
	\begin{center}	
		\caption{\label{tab:1}The properties of the states $K^*$~\cite{ParticleDataGroup:2022pth}. The mass and width values are in units of MeV.}
		\begin{tabular}{cccccc}
			\hline\hline
			states     & $J^{P}$  & mass  & width &  $\mathcal{B}$($K^*\to K\pi$) & $\mathcal{B}$($B_s\to \bar{D}^0\bar{K}^{0*}$)  \\
			\hline
			$K^*(892)$  &  $1^-$ &895.55  &47.3  &100$\%$  &$4.4\times10^{-4}$ \\
			
			$K^*(1410)$  &  $1^-$ &1414  &232  &6.6$\%$  &$3.9\times10^{-4}$ \\
			
			$K_0^*(1430)$  &  $0^+$ &1425  &270  &93$\%$  &$3.0\times10^{-4}$ \\
			
			$K_2^*(1430)$  &  $2^+$ &1427.3  &100  &49.9$\%$  &$1.1\times10^{-4}$ \\
			
			\hline\hline
		\end{tabular}
	\end{center}
\end{table}

In addition to the $T_{c\bar{s}}(2900)$, there could be some contributions from the excited kaons. As shown in Table~\ref{tab:1}, one can find that, $\mathcal{B}(B_s\to \bar{D}^0\bar{K}^*(892))=(4.4\pm 
0.6)\times 10^{-4}$, $\mathcal{B}(B_s\to \bar{D}^0\bar{K}
^*(1410))=(3.9\pm 3.5)\times 10^{-4}$, $\mathcal{B}(B_s\to \bar{D}^0\bar{K}_0^*(1430))=(3.0\pm 0.7)\times 10^{-4}$, and $\mathcal{B}(B_s\to \bar{D}^0\bar{K}_2^*(1430))=(1.1\pm 0.4)\times 10^{-4}$,
which implies that $K^*(892)$, $K^*(1410)$, $K_0^*(1430)$, and $K_2^*(1430)$ could play important roles in the process $\bar{B}
_s^0\to  D^0K^0\pi^0$.  However, since $\mathcal{B}(B_s\to \bar{D}^0\bar{K}
^*(1410))=(3.9\pm 3.5)\times 10^{-4}$ has large uncertainties and the branching fraction $\mathcal{B}(K^*(1410)\to K \pi)=(6.6\pm1.3)\%$ is much smaller than those of the other three excited kaons~\cite{ParticleDataGroup:2022pth}, as shown in Table~\ref{tab:1}, we neglect its contribution and only take into account the contributions of $K^*(892)$, $K_0^*(1430)$, and $K_2^*(1430)$ in this work. 

For the process $\bar{B}_s^0\to D^0K^*(892) \to D^0K^0\pi^0$  depicted in Fig.~\ref{fig:Kstar}, the amplitude can be written as
\begin{equation}\label{t892}
	\mathcal{T}^{K^*(892)} =\dfrac{V^{K^*(892)}|\vec{p}^{\,*}_{\pi^0}||\vec{p}^{\,*}_{D^0}|{\rm cos}\theta}{M^2_{K^0\pi^0}-M^2_{K^*(892)}+iM_{K^*(892)}\Gamma_{K^*(892)}}  ,
\end{equation}
where $V^{K^*(892)}$ is the relative strength of the contribution from the intermediate resonance $K^*(892)$. We can calculate the value utilizing the branching fraction $\mathcal{B}(\bar{B}_s^0\to D^0K^*(892))\times\mathcal{B}(K^*(892)\to K^0\pi^0)=4.4\times10^{-4}\times100\%\times1/3$~\cite{LHCb:2014ioa,ParticleDataGroup:2022pth}, and roughly estimate $V^{K^*(892)}=1.159\times10^{-7}$. $|\vec{p}^{\,*}_{\pi}|$ and $|\vec{p}^{\,*}_{D}|$ are the momenta of $\pi^0$ and $D^0$ in the rest frame of the $K^0\pi^0$ system, respectively, and $\theta$ is the angle between $\pi^0$ and $D^0$ in the center of mass frame of the $K^0\pi^0$ system, as depicted in Fig.~\ref{fig:angle-diagram}, which are given by~\cite{Wang:2015pcn,Wang:2022nac,Lyu:2024qgc},
\begin{equation}\label{ppi}
	|\vec{p}^{\,*}_{\pi^0}|=\dfrac{\lambda^{1/2}(M_{K^0\pi^0}^2,m_{\pi^0}^2,M_{K^0}^2)}{2M_{K^0\pi^0}},
\end{equation}
\begin{equation}\label{pd}
	|\vec{p}^{\,*}_{D^0}|=\dfrac{\lambda^{1/2}(M_{\bar{B}_s^0}^2,m_{D^0}^2,M_{K^0\pi^0}^2)}{2M_{K^0\pi^0}},
\end{equation}
\begin{equation}\label{theta}
	{\rm cos}\theta=\dfrac{M_{D^0K^0}^2-M_{\bar{B}_s^0}^2-m_{\pi^0}^2+2P_{\bar{B}_s^0}^0P_{\pi^0}^0}{2|\vec{p}_{\pi^0}||\vec{p}_{D^0}|},
\end{equation}
with the K$\ddot{a}$llen function $\lambda(x,y,z)=x^2+y^2+z^2-2xy-2yz-2xz$.
In the $K^0\pi^0$ rest frame, $\vec{p}_{\bar{B}_s^0}=\vec{p}_{D^0}$, $\vec{p}_{\pi^0}=-\vec{p}_{K^0}$, and the $\bar{B}_s^0$ and $\pi^0$ energies are,
\begin{equation}
	\begin{aligned}
		P_{\bar{B}_s^0}^0&=\sqrt{M_{\bar{B}_s^0}^2+|\vec{p}_{\bar{B}_s^0}|^2}=\sqrt{M_{\bar{B}_s^0}^2+|\vec{p}_{D^0}|^2}, \nonumber\\
		P_{\pi^0}^0&=\sqrt{m_{\pi^0}^2+|\vec{p}_{\pi^0}|^2}.
	\end{aligned}
\end{equation}

We will also take into account the contribution of the resonance $K_0^*(1430)$ in the process $\bar{B}_s^0\to D^0K_0^*(1430)\to D^0K^0\pi^0$, as shown in Fig.~\ref{fig:Kstar}, of which the amplitude can be written as 
\begin{equation}\label{t1430b}
	\mathcal{T}^{K_0^*(1430)} =\dfrac{V^{K_0^*(1430)}M_{K_0^*(1430)}\Gamma_{K_0^*(1430)}}{M^2_{K^0\pi^0}-M^2_{K_0^*(1430)}+iM_{K_0^*(1430)} \Gamma_{K_0^*(1430)}}  ,
\end{equation}
where $V^{K_0^*(1430)}$ is the relative strength of the contribution from the intermediate resonance $K_0^*(1430)$. We can calculate the value using the branching fraction $\mathcal{B}(\bar{B}_s^0\to D^0K_0^*(1430))\times\mathcal{B}(K_0^*(1430)\to K^0\pi^0)=3.0\times10^{-4}\times93\%\times1/3$~\cite{LHCb:2014ioa,ParticleDataGroup:2022pth}, and roughly estimate $V^{K_0^*(1430)}=1.533\times10^{-6}$.

As we discussed in the introduction, in addition to $K^*(892)$ and $K_0^*(1430)$, we will also take into account the contribution of the resonance $K_2^*(1430)$ in the process $\bar{B}_s^0\to D^0K_2^*(1430)\to D^0K^0\pi^0$, as shown in Fig.~\ref{fig:Kstar}.
The amplitude of Fig.~\ref{fig:Kstar} considering the contribution of $K_2^*(1430)$ can be written as~\cite{Zhang:2022xpf,Wang:2015pcn},
\begin{equation}\label{t1430a}
	\mathcal{T}^{K_2^*(1430)} =\dfrac{V^{K_2^*(1430)}|\vec{p}^{\,*}_{\pi^0}|^2({3\rm cos}^2\theta-1)}{M^2_{K^0\pi^0}-M^2_{K_2^*(1430)}+iM_{K_2^*(1430)} \Gamma_{K_2^*(1430)}}  ,
\end{equation}
where $V^{K_2^*(1430)}$ is the relative strength of the contribution from the intermediate resonance $K_2^*(1430)$. We can calculate the value utilizing the branching fraction $\mathcal{B}(\bar{B}_s^0\to D^0K_2^*(1430))\times\mathcal{B}(K_2^*(1430)\to K^0\pi^0)=1.1\times10^{-4}\times49.9\%\times1/3$~\cite{LHCb:2014ioa,ParticleDataGroup:2022pth}, and roughly estimate $V^{K_2^*(1430)}=3.996\times10^{-7}$.

In this work, we take the resonance masses and widths as $M_{K^*(892)}=895.55$~MeV and $\Gamma_{K^*(892)}=47.3$~MeV, $M_{K_0^*(1430)}=1425$~MeV and $\Gamma_{K_0^*(1430)}=270$~MeV, and $M_{K_2^*(1430)}=1427.3$~MeV and $\Gamma_{K_2^*(1430)}=100$~MeV from Ref.~\cite{ParticleDataGroup:2022pth}.
	
\subsection{Invariant Mass Distribution} \label{sec2e}
	
	\begin{figure}
		\subfigure[]{
			\includegraphics[scale=0.65]{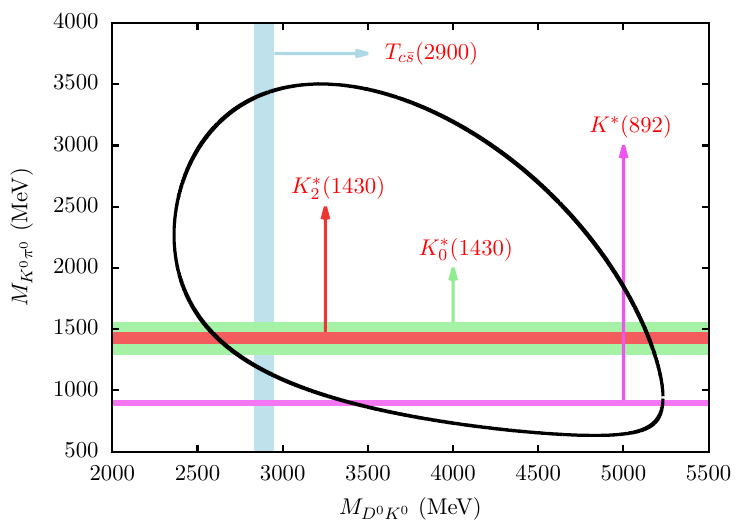}
		}
		\subfigure[]{
			\includegraphics[scale=0.65]{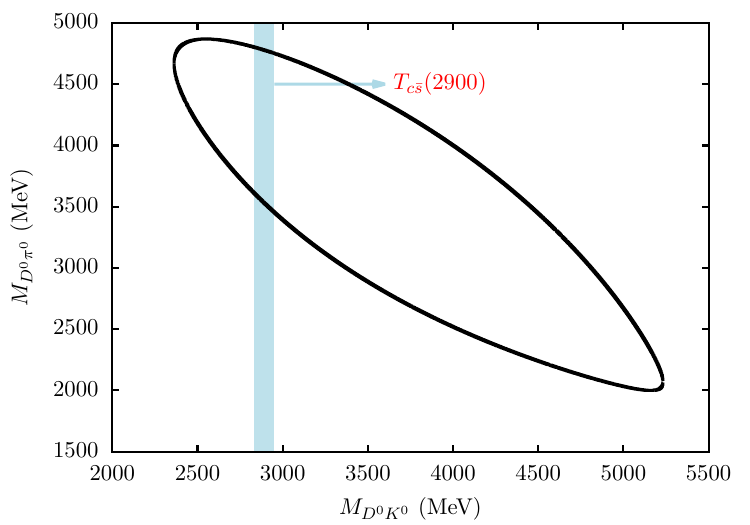}
		}
		\caption{The Dalitz plots for the process $\bar{B}_s^0\to K^0D^0\pi^0$. The green band stands for the region of 1290$\sim$1560~MeV that the predicted $K_0^*(1430)$ state lies in, the blue band stands for the region of 2833$\sim$2952~MeV that the $T_{c\bar{s}}(2900)$ state lies in, the magenta band stands for the region of 872$\sim$920~MeV that the $K^*(892)$ state lies in, the red band stands for the region of 1377$\sim$1477~MeV that the $K_2^*(1430)$ state lies in.}\label{fig:Dalitz}
	\end{figure}
	
With all the ingredients obtained in the previous sections, one can write down the double differential decay width for the process $\bar{B}_s^0\to K^0D^0\pi^0$ as follows,
\begin{equation}\label{eq:dwidth}
	\dfrac{d^2\Gamma}{dM_{K^0\pi^0}dM_{D^0K^0}} = \frac{1}{(2\pi)^3}\dfrac{2M_{K^0\pi^0}2M_{D^0K^0}}{32M_{\bar{B}_s^0}^3}|\mathcal{T}^{\text {Total}}|^2,
\end{equation}
where the total amplitude is,
\begin{equation}\label{eq:totalamp}
		\begin{aligned}
	|\mathcal{T}^{\text {Total}}|^2
	=&|\mathcal{T}^{BG}+\mathcal{T}^{T_{c\bar{s}}}+\mathcal{T}^{K_0^*(1430)}+\mathcal{T}^{K^*(892)} \\
     &+\mathcal{T}^{K_2^*(1430)}|^2.
     \end{aligned}
\end{equation}
where $\mathcal{T}^{BG}$ is the contribution from the tree diagram of $\bar{B}^0_s\to K^0 D^0 \pi^0$, similar as Fig.~\ref{fig:Tcs-quark}(b). Thus, we assume the same weight as the one of Fig.~\ref{fig:Tcs-quark}(b),  
\begin{equation}
	\mathcal{T}^{BG}=-\frac{1}{\sqrt{2}} V_p.
\end{equation}

The invariant mass distributions $d\Gamma/dM_{K^0\pi^0}$ and $d\Gamma/dM_{D^0K^0}$ can be obtained by integrating Eq.~(\ref{eq:dwidth}) over each of the invariant mass variables. For a given value of $M_{12}$, the range of $M_{23}$ could be determined~\cite{ParticleDataGroup:2022pth} by
	\begin{align}
		&\left(m_{23}^2\right)_{\min}=\left(E_2^*+E_3^*\right)^2-\left(\sqrt{E_2^{* 2}-m_2^2}+\sqrt{E_3^{* 2}-m_3^2}\right)^2, \nonumber\\
		&\left(m_{23}^2\right)_{\max}=\left(E_2^*+E_3^*\right)^2-\left(\sqrt{E_2^{* 2}-m_2^2}-\sqrt{E_3^{* 2}-m_3^2}\right)^2, 
	\end{align}
	where $E_2^{*}$ and $E_3^{*}$ are the energies of particles 2 and 3 in the $M_{12}$ rest frame, which are written as,
	\begin{align}
		&E_2^{*}=\dfrac{M_{12}^2-m_1^2+m_2^2}{2M_{12}}, \nonumber\\
		&E_3^{*}=\dfrac{M_{\Lambda_b}^2-M_{12}^2+m_3^2}{2M_{12}},
	\end{align}
	with $m_1$, $m_2$, and $m_3$ are the masses of involved particles 1, 2, and 3, respectively. 
All masses and widths of the particles are taken from the Review of Particle Physics (RPP)~\cite{ParticleDataGroup:2022pth}.
	
On the other hand, the process $\bar{B}_s^0\to K^0D^0\pi^0$ can also occur through the intermediate states $D^*$ decay into $D^0\pi^0$. We can easily understand from the Dalitz plots in Fig.~\ref{fig:Dalitz}(b) that the $T_{c\bar{s}}(2900)$ position is in the high-energy region of the $D^0\pi^0$ invariant mass distribution. 

It should be stressed that $\mathcal{B}(B_s\to \bar{D}^* \bar{K}^0)=(2.8\pm 1.1)\times 10^{-4}$~\cite{ParticleDataGroup:2022pth,LHCb:2016lky}. Considering that the width of the state $D^0$ is very small ($<2.1$~MeV)~\cite{ParticleDataGroup:2022pth}, and the contributions of the $D^*$ and also other excited $D$ mesons hardly affect the signal of $T_{c\bar{s}}(2900)$, as shown by Fig.~\ref{fig:Dalitz}(b). Thus, we neglect the possible contribution from intermediate excited $D$ mesons.

\section{Results and Discussions}\label{sec3}

\begin{figure}[htbp]
	\subfigure{
		\includegraphics[scale=0.65]{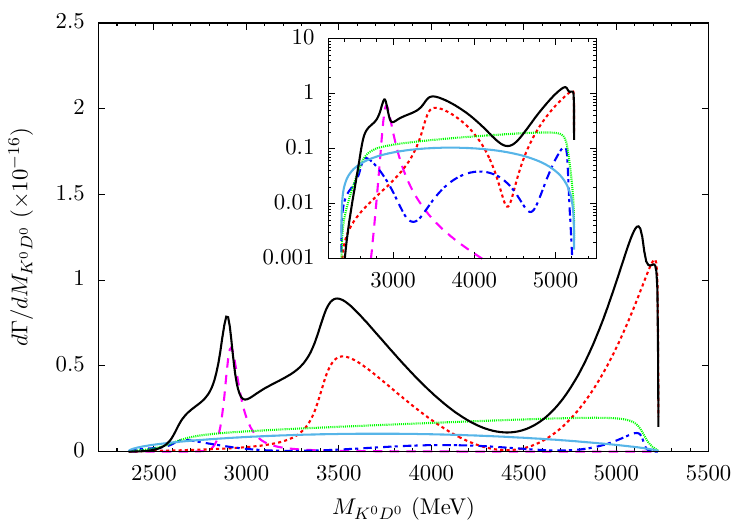}
	}
	\subfigure{
		\includegraphics[scale=0.65]{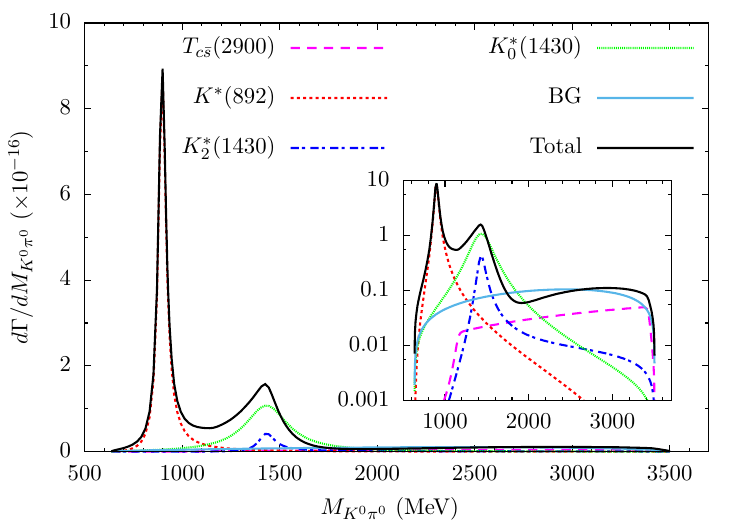}
	}
	\caption{The $D^0K^0$ and $K^0\pi^0$ invariant mass distributions of the process $\bar{B}_s^0\to K^0D^0\pi^0$ decay. The inner sub-figures show the results with the vertical axis represented in logarithmic scale.   }\label{fig:dwidth}
\end{figure}

\begin{figure}
	\centering
	\includegraphics[scale=0.80]{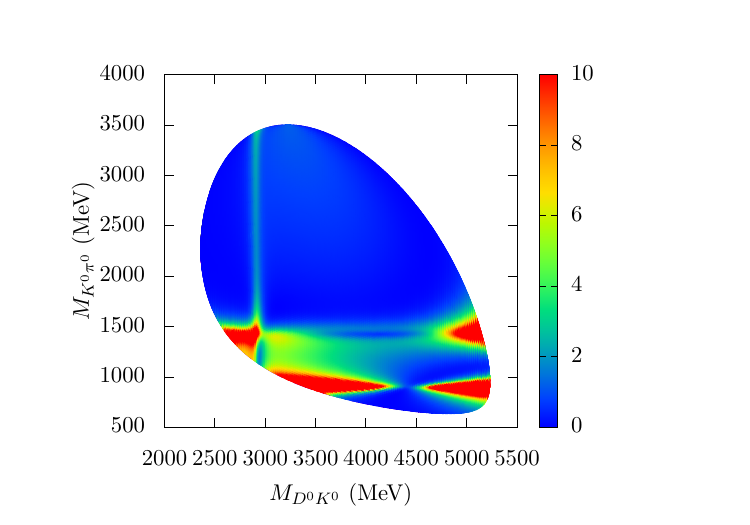}
	\caption{The Dalitz plot of ``$M_{D^0K^0}$" vs. ``$M_{K^0\pi^0}$" for the double differential decay width $d^2\Gamma/dM_{D^0K^0}dM_{K^0\pi^0}$ the process $\bar{B}_s^0\to K^0D^0\pi^0$. }\label{fig:Dalize-DK-Kpi}
\end{figure}
\begin{figure}
	\centering
	\includegraphics[scale=0.65]{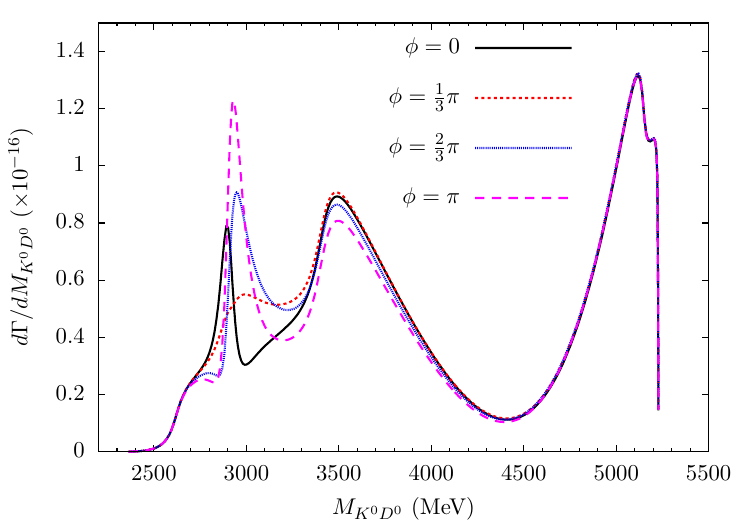}
	\caption{The $D^0K^0$ invariant mass distribution of the process $\bar{B}_s^0\to K^0D^0\pi^0$ decay with the interference phase $\phi=0$, ${\pi}/{3}$, ${2\pi}/{3}$, and $\pi$.}\label{fig:dwidth-DK-phi}
\end{figure}
\begin{figure}
	\centering
	\includegraphics[scale=0.65]{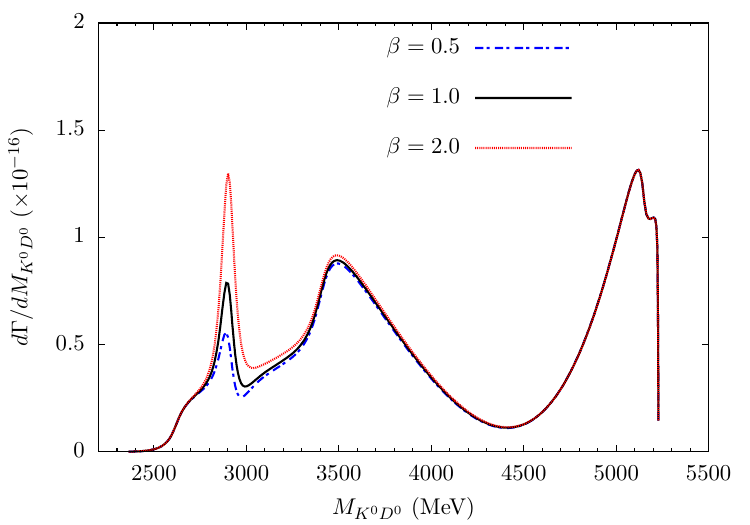}
	\caption{The $D^0K^0$ invariant mass distribution of the process $\bar{B}_s^0\to K^0D^0\pi^0$ decay with the ration $\beta=0.5$, $1.0$, and $2.0$.}\label{fig:dwidth-DK-beta}
\end{figure}
	
\begin{figure}[htbp]
	\centering
	\includegraphics[scale=0.65]{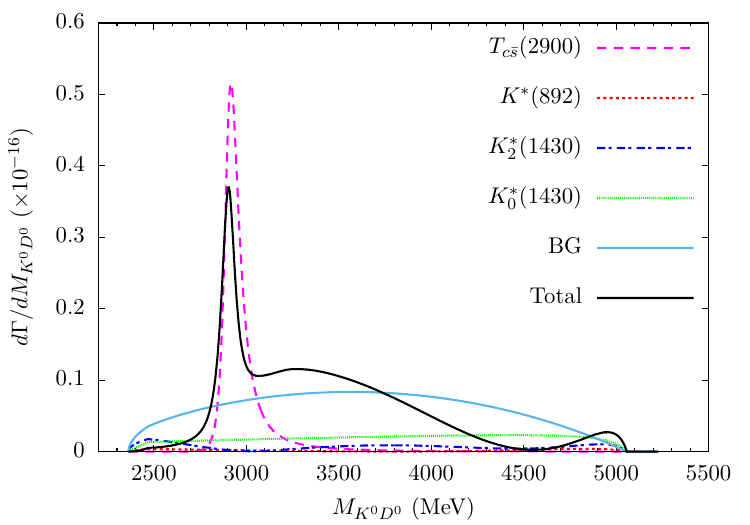}
	\caption{The $D^0K^0$ invariant mass distribution of the process $\bar{B}_s^0\to K^0D^0\pi^0$ decay restricted to $M_{K^0\pi^0}\geq1700$~MeV.}\label{fig:dwidth-DK-cut}
\end{figure}

In our model, there is only one free parameter $V_{p}$, corresponding to the strength of the contribution of $T_{c\bar{s}}(2900)$. Currently, there is no experimental information that could be used to extract the parameter $V_p$. In our calculation, since the branching fraction $\mathcal{B}(\bar{B}_s^0\to D^0K_2^*(1430)\to D^0K^0\pi^0)$ is listed, we assume that $\mathcal{B}(\bar{B}_s^0\to T_{c\bar{s}}(2900)\pi^0\to D^0K^0\pi^0)$ equals this branching fraction to calculate $V_{p}$. And we obtain the value $V_{p}=5.678\times10^{-7}$, and we will discuss the uncertainties from the different ratio $\beta=\mathcal{B}(\bar{B}_s^0\to T_{c\bar{s}}(2900)\pi^0\to D^0K^0\pi^0)/\mathcal{B}(\bar{B}_s^0\to D^0K_2^*(1430)\to D^0K^0\pi^0)$ later.

We present our results of the $D^0K^0$ and $K^0\pi^0$ invariant mass distributions, as depicted in Fig.~\ref{fig:dwidth}. The magenta-dashed curves show the contribution of $\mathcal{T}^{T_{c\bar{s}}}$, the red-dotted curves show the contribution of $\mathcal{T}^{K^*(892)}$, the blue-dashed-dotted curves show the contribution from $\mathcal{T}^{K_2^*(1430)}$, the green-dotted curves show the contribution from $\mathcal{T}^{K_0^*(1430)}$, the blue-solid curves show the contribution from $\mathcal{T}^{BG}$, and the black-solid curves show the contribution from the total amplitude. One can find a significant peak around 2900~MeV in the $D^0K^0$ invariant mass distribution, which could be associated with the tetraquark state $T_{c\bar{s}}(2900)$. Meanwhile, one can find a peak around 900~MeV in the $K^0\pi^0$ invariant mass distribution, which is due to the resonance $K^*(892)$, and a peak around 1430~MeV in the $K^0\pi^0$ invariant mass distribution, which is due to the resonances $K_0^*(1430)$ and $K_2^*(1430)$.

Next, we present the Dalitz plot for the process $\bar{B}_s^0\to K^0D^0\pi^0$ depending on $M_{D^0K^0}$ and $M_{K^0\pi^0}$ in Fig.~\ref{fig:Dalize-DK-Kpi}. One can find that $K^*(892)$ mainly contributes to the high-energy region of the $D^0K^0$ invariant mass distribution, and a clear band, corresponding to $T_{c\bar{s}}(2900)$, appears in the region around $M_{D^0K^0}=2900$~MeV. And one can see a clear interference effect from $T_{c\bar{s}}(2900)$ and $K_0^*(1430)$ in $M_{D^0K^0}=2900$~MeV and $M_{K^0\pi^0}=1430$~MeV.

In addition, it should be noted that there may be phase interference between $\mathcal{T}^{T_{c\bar{s}}}$ and others in Eq.~(\ref{eq:totalamp}). Thus, we multiply the amplitude $\mathcal{T}^{T_{c\bar{s}}}$ by a phase factor $e^{i\phi}$, and calculate the $D^0K^0$ invariant mass distribution with $\phi=0$, $\pi/3$, ${2\pi}/{3}$, and $\pi$, as presented in Fig.~\ref{fig:dwidth-DK-phi}. Although the lineshape of the peak structure in the $D^0K^0$ invariant mass distribution is distorted by interference with different phase angles $\phi$, one can always find a clear signal of the peak structure around 2900~MeV.

In addition, we have assumed the relationship $\mathcal{B}(\bar{B}_s^0\to T_{c\bar{s}}(2900)\pi^0\to D^0K^0\pi^0)=\mathcal{B}(\bar{B}_s^0\to D^0K_2^*(1430)\to D^0K^0\pi^0)$ to obtain the parameters $V_p$ in our calculations. To check the uncertainty of the weight parameter $V_p$\footnote{One can easily find that the uncertainty of $V_p$ is also related to the uncertainty of the width $\Gamma_{T_{c\bar{s}}\to DK}$.}, we have also presented the $D^0K^0$ invariant mass distribution with different values of $\beta=\mathcal{B}(\bar{B}_s^0\to T_{c\bar{s}}(2900)\pi^0\to D^0K^0\pi^0)/\mathcal{B}(\bar{B}_s^0\to D^0K_2^*(1430)\to D^0K^0\pi^0)=0.5$, $1.0$, and $2.0$ in Fig.~\ref{fig:dwidth-DK-beta}. One can find that the peak of $T_{c\bar{s}}(2900)$ in the $D^0K^0$ invariant mass distribution is always clear.

Then we find that the resonance $K^*(892)$ does not affect the structure of $T_{c\bar{s}}(2900)$, which can be easily understood from the Dalitz plots of Figs.~\ref{fig:Dalitz}(a) and \ref{fig:Dalize-DK-Kpi}. Thus, we take the cut of $M_{K^0\pi^0}\geq1700$~MeV to eliminate the contributions from the $K^*(892)$, $K^*_0(1430)$, and $K^*_2(1430)$, and show the results in Fig.~\ref{fig:dwidth-DK-cut}. One can find that the structure of $T_{c\bar{s}}(2900)$ is not affected, while the signal becomes clearer.

\section{ Conclusions }
	
Motivated by the LHCb Collaboration measurements of the decays $B^0\to\bar{D}^0D_s^+\pi^-$ and $B^+\to D^-D_s^+\pi^+$, we have investigated the process $\bar{B}_s^0\to K^0D^0\pi^0$ by taking into account the $D_s^{*+}\rho^-$ and $D^{*0}K^{*0}$ final-state interaction to generate the tetraquark state $T_{c\bar{s}}(2900)$. In addition, we have also taken into account the contributions from the resonances $K^*(892)$, $K_0^*(1430)$, and $K_2^*(1430)$.
	
Our results show that a significant peak structure could be found around 2900~MeV in the $D^0K^0$ invariant mass distribution, which could be associated with the open-flavored tetraquark state $T_{c\bar{s}}(2900)$. It should be pointed out that there are some unknown parameters, and we have discussed the uncertainties resulting from those parameters. Future precise measurements of the processes $\bar{B}_s^0\to K^0D^0\pi^0$ could be used to constrain those parameters.
	
Since both the tree-diagram amplitudes of processes $\bar{B}_s^0\to K^0D^0\pi^0$ and $B_s^0\to \bar{D}^0K^-\pi^+$ are proportional to the CKM matrix elements $V_{cb}V_{ud}$, it is expected that the branching fractions of those two processes should be of the same order of magnitude, i.e. $10^{-3}$, which implies that the process $\bar{B}_s^0\to K^0D^0\pi^0$ could be accessible to be measured by LHCb in the future. 
If the proposed process is measured in the future, one can obtain more precise properties about the $T_{c\bar{s}}(2900)$, such as the branching fraction of the $T_{c\bar{s}}(2900)\to K^0D^0$, which should be helpful to distinguish its different explanations. 
On the other hand, for the $T_{c\bar{s}}(2900)^{++}$ with the quark contents $c\bar{s}\bar{d}u$, one could search for it in the $D_s^+ \pi^+$ or $D^+K^+$ mode, such as the processes $B^+\to D^+K^+\pi^-$, $B^+\to D_s^+\pi^+\pi^-$, $B^+\to D^+D^-K^+$, $B^+\to D_s^+D_s^-\pi^+$, $\bar{B}_s^0\to D_s^+\pi^-\pi^0$ and $B_c^+\to D_s^+D_s^-\pi^+$. Therefore, we strongly encourage our experimental colleagues to measure these processes, which would be crucial to confirm the existence of the $T_{c\bar{s}}(2900)$ resonance.

\section*{Acknowledgements}
We would like to acknowledge the fruitful discussions with Prof. Eulogio Oset. 
E.Wang acknowledge the support from the National Key R\&D Program of China (No. 2024YFE0105200).	
This work is supported by the Natural Science Foundation of Henan under Grant No. 232300421140, No. 242300421377 and No. 222300420554, the National Natural Science Foundation of China under Grant No. 12475086, No. 12192263, No. 12175037, and No. 12335001.

\end{document}